\documentclass{PoS}

\newcommand{\as}{\alpha_\mathrm{s}}

\title{TMDs and Monte Carlo Event Generators}

\ShortTitle{TMDs and MC Event Generators}

\author{\speaker{F Hautmann}
\\
  University of Oxford / University of Antwerp\\
  Also at:  UPV/EHU, Bilbao and CERN, Geneva\\       
        E-mail: \email{francesco.hautmann@physics.ox.ac.uk}}

\abstract{We discuss 
prospects for Monte Carlo  event generators  
incorporating the dynamics of
  transverse momentum dependent (TMD) parton 
distribution functions. We  illustrate TMD evolution 
 in the parton branching formalism, and present Monte Carlo applications of the method.}

\FullConference{23rd International Spin Physics Symposium - SPIN2018 -\\
		10-14 September, 2018\\
		Ferrara, Italy}

\begin{document}

\section{Introduction}

Transverse momentum dependent parton 
distribution functions (TMD PDFs, or TMDs for short) 
encode nonperturbative information on hadron 
structure, including transverse momentum and polarization degrees of freedom, and enter QCD factorization 
theorems  for physical observables in hadronic collisions 
both in the Sudakov region~\cite{jcc-book,css85} 
and in the high-energy region~\cite{cch91,ch94}. 
They  provide a ``3-dimensional imaging'' of 
hadron structure, extending to the transverse plane 
the 1-dimensional picture given by  collinear 
PDFs~\cite{Kovarik:2019xvh}. 

 While a great amount of knowledge has been built about 
collinear PDFs from  experiments in hadron collisions over the past thirty years, 
TMDs are much less known,  and hadronic 3-D imaging, with  its 
implications for high-energy physics, will constitute the subject of 
intensive studies in the forthcoming decade.  
 Experimental data analyses 
require 
 realistic  Monte Carlo  event simulations.   
 The construction of Monte Carlo 
 generators incorporating 
TMDs 
is 
thus 
a central objective of this physics program. 

This article  describes  progress  in this direction based on the works~\cite{plb17soft,jhep18}. 
 Sec.~2  discusses  the  parton branching (PB)  formulation of TMD evolution.   Sec.~3   presents Monte Carlo (MC) calculations using 
PB TMDs with  applications to deep inelastic scattering (DIS) and Drell-Yan (DY) processes.   Sec.~4 gives concluding remarks.

\section{TMD evolution in the parton branching formalism}

Table~1 gives the full set of quark (left) and gluon (right) unpolarized and polarized TMD distributions in a spin-1/2 hadron~\cite{Angeles-Martinez:2015sea}. 
Columns represent parton polarization, 
rows represent  hadron polarization. 
The $f$, $g$ and $h$ distributions on the diagonal are respectively the unpolarized,  helicity 
and  transversity TMDs.
The blue and pink shades of the boxes 
indicate respectively T-even and T-odd distributions,
 i.e., involving an even or odd number of  spin flips. 
In this article we   concentrate   on   
QCD evolution in the  unpolarized case. 

Although TMDs are nonperturbative quantities, QCD factorization theorems, 
combined with renormalization group analysis, imply 
that the evolution of TMDs with 
mass and/or energy scales can be expressed in terms of perturbative kernels, computable as 
power series expansions in the QCD coupling $\as$. 
Well-known examples  are 
provided  by the CSS evolution 
equation (or its variants) in the Sudakov region, and 
BFKL evolution equation (or its variants) in the 
high-energy region. These 
achieve the resummation of 
logarithmically-enhanced 
radiative corrections (double-log in 
the Sudakov case, single-log in the high-energy case) 
to all orders in  $\as$, and give rise to a successful 
phenomenology, even though limited in each case 
to a specific kinematic region and a  
specific  set of  observables.  

In Refs.~\cite{plb17soft,jhep18} a different approach from the  above formalisms   
  is proposed, which   
 looks  for a  formulation of TMD evolution  that  can be useful in  more general 
collider kinematics and for 
 broader classes of observables,  aiming to  
 fulfill  the 
following criteria: i) applicability over a wide kinematic range from low to high transverse 
momenta; ii) implementability in Monte Carlo event 
generators describing  the exclusive structure of the 
 final states; iii)  evolution equations which are 
connected in a controllable way with DGLAP evolution 
equation of collinear PDFs. To this end, Refs.~\cite{plb17soft,jhep18}  use the unitarity picture of 
parton evolution~\cite{Webber:1986mc,eswbook} (commonly employed in parton shower algorithms).   
Soft gluon emission and transverse momentum recoils are treated 
by introducing the soft-gluon resolution scale to separate 
resolvable and non-resolvable branchings  and the Sudakov form factor to express  the probability for no resolvable  branching 
in a given evolution interval. The TMD evolution equation in the  parton branching (PB) approach is given by~\cite{jhep18}

\begin{table}
\centering
\includegraphics[scale=0.3]{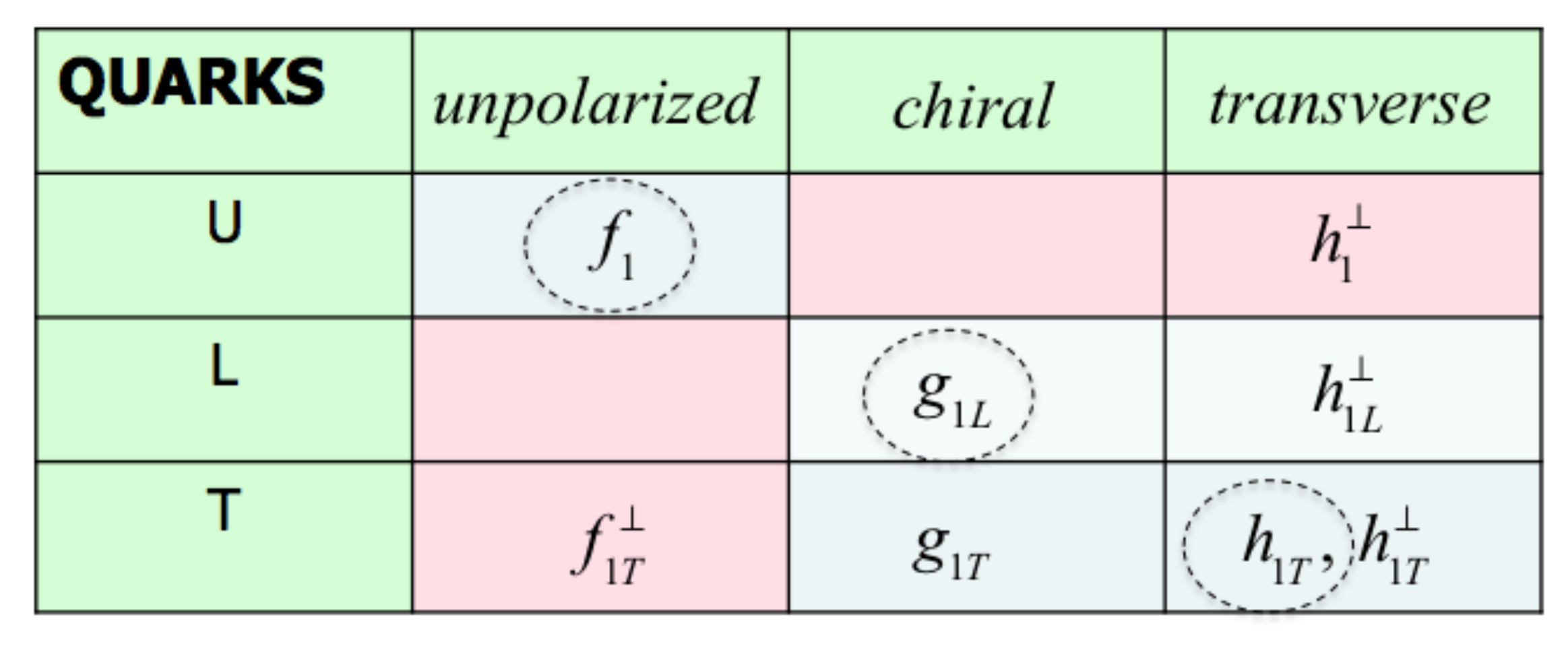}
\includegraphics[scale=0.3]{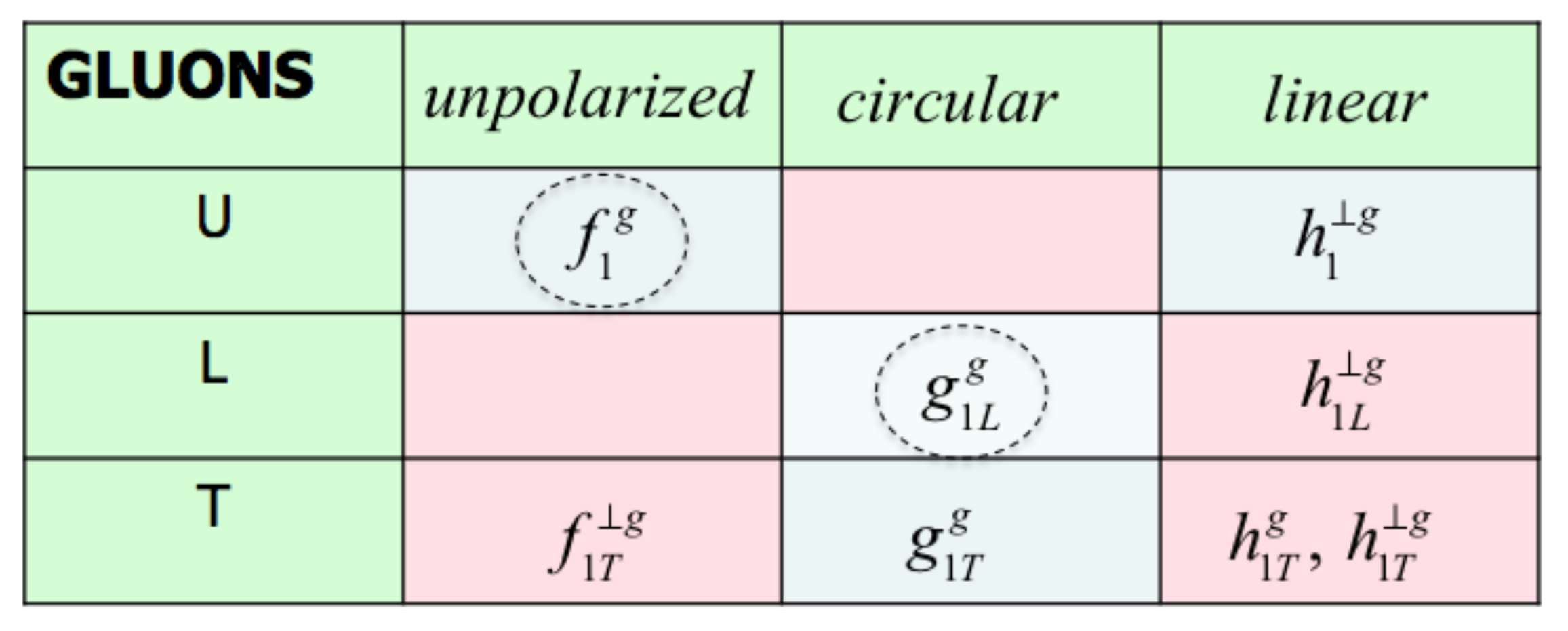}
\caption{\it  (left) Quark and (right) gluon 
TMD PDFs~\protect\cite{Angeles-Martinez:2015sea}. 
}
\label{t:gluon_tw2_tmdpdfs}
\end{table}

\begin{eqnarray}
\label{integeqforA}
  { {\cal A}}_a(x,{\bf k}, \mu^2) 
 &=&  
\Delta_a (  \mu^2  ) \ 
 { {\cal A}}_a(x,{\bf k},\mu^2_0)  
 + \sum_b 
\int
{{d^2 {\bf q}^{\prime } } 
\over {\pi {\bf q}^{\prime 2} } }
 \ 
{
{\Delta_a (  \mu^2  )} 
 \over 
{\Delta_a (  {\bf q}^{\prime 2}  
 ) }
}
\ \Theta(\mu^2-{\bf q}^{\prime 2}) \  
\Theta({\bf q}^{\prime 2} - \mu^2_0)
 \nonumber\\ 
&\times&  
\int_x^{z_M} {{dz} \over z}  \;
P_{ab}^{(R)} (\as,z) 
\;{ {\cal A}}_b({x/z}, {\bf k}+(1-z) {\bf q}^\prime , 
{\bf q}^{\prime 2})  
  \;\;   , 
\end{eqnarray}
where ${ {\cal A}}_a(x,{\bf k}, \mu^2) $ is the TMD distribution of flavor $a$ at longitudinal momentum fraction $x$, 
transverse momentum ${\bf k}$, evolution scale $\mu$;   $z_M$ is the soft-gluon resolution scale; $ P_{ab}^{(R)}$ are the real-emission 
splitting kernels, computed as a perturbative series expansion in  $\as$ to leading order (LO), next-to-leading-order (NLO), etc.; 
 $\Delta_a$ is the  Sudakov form factor, given by 
\begin{equation}
\label{sud-def}
 \Delta_a (  \mu^2  ) = 
\exp \left(  -  \sum_b  
\int^{\mu^2}_{\mu^2_0} 
{{d \mu^{\prime 2} } 
\over \mu^{\prime 2} } 
 \int_0^{z_M} dz \  z 
\ P_{ba}^{(R)}(\as , 
 z ) 
\right) 
  \;\; .  
\end{equation}

An important point in obtaining TMD distributions from the PB method 
concerns the ordering variables used to perform the branching evolution. 
The basic issue is that the transverse momentum 
generated radiatively 
by the recoils 
in the  evolution cascade depends 
 on the treatment of the 
non-resolvable region $z \to 1$~\cite{Hautmann:2007uw}.   
While in the collinear 
distribution, obtained by integration over  ${\bf k }$,  
 $z \to 1$ singularities  cancel 
between real and virtual 
non-resolvable emissions, 
this is not guaranteed in 
 the case of the TMD distribution, and 
 supplementary conditions are needed. 
In the  PB method  these are provided by gluon emissions' angular ordering. 
The rescaling and shift in the transverse momentum argument of 
${ {\cal A}}$ in the last term on the r.h.s.~of   Eq.~(\ref{integeqforA}) take  
into account the angular ordering condition. 

By integrating Eq.~(\ref{integeqforA}) over transverse momenta  one obtains collinear initial-state distributions.  
For $z_M \to 1$, these are    collinear PDFs, and one recovers DGLAP evolution equations.  
The convergence to DGLAP has been verified numerically 
in~\cite{jhep18}  at NLO  to better than 1 \%  over several orders of magnitude  in $x$ and $\mu$.   This is in a similar  spirit to the 
earlier studies~\cite{Jadach:2003bu,tanaka03}.    
For general $z_M$, the evolution equation that follows from integrating  
Eq.~(\ref{integeqforA}) coincides with the coherent  branching equation~\cite{Catani:1990rr,Marchesini:1987cf}. 

The TMD branching equation (\ref{integeqforA}) can be solved by Monte Carlo methods. Such a 
 solution has been presented in~\cite{jhep18}.  This allows one not only to  determine  the inclusive distributions but also to 
 reconstruct exclusively  the radiative final states. 
The principle on which the PB method is based is similar to that of parton showers, but the difference is that in the PB method   
nonperturbative TMD densities are defined and determined from fits to experimental data, which places  constraints on fixed-scale 
inputs to evolution.  This is in the spirit of approaches discussed e.g. in~\cite{tmdplott,jccandted15}, and  is in contrast to 
parton shower  Monte Carlo calculations,  in which 
parton densities are not used to  constrain evolution,  while  nonperturbative physics parameters  are tuned.  
Monte Carlo applications of the PB method are presented in the next section for deep inelastic scattering (DIS) and Drell-Yan (DY) 
 lepton pair production.

\section{Inclusive DIS and DY $p_T$ distribution}

Applications of the PB approach to physical observables in hadronic collisions require the use of perturbative matrix elements for 
the production of a  hard final state with high momentum transfer along with the multi-parton cascade from the TMD branching equation 
(\ref{integeqforA}). 
The distributions $  { {\cal A}}_a(x,{\bf k},\mu^2_0) $ in the first term on the r.h.s.~of 
Eq.~(\ref{integeqforA})    are  initial conditions   
which represent the intrinsic $x$ and $k_T$ distributions at scale $\mu_0$, and are  
to be determined from fits to experiment. 
In this section we describe the results of determining these distributions  
at NLO from  the high-precision  inclusive  DIS measurements,  and using  
the branching equation (\ref{integeqforA})  along with NLO  perturbative matrix elements for $Z$-boson DY hadroproduction 
to obtain the PB - TMD predictions for  DY transverse momentum $p_T$ and $\phi^*$  spectra.  

\begin{figure}[h!tb]
\begin{center} 
\includegraphics[width=0.3\textwidth]{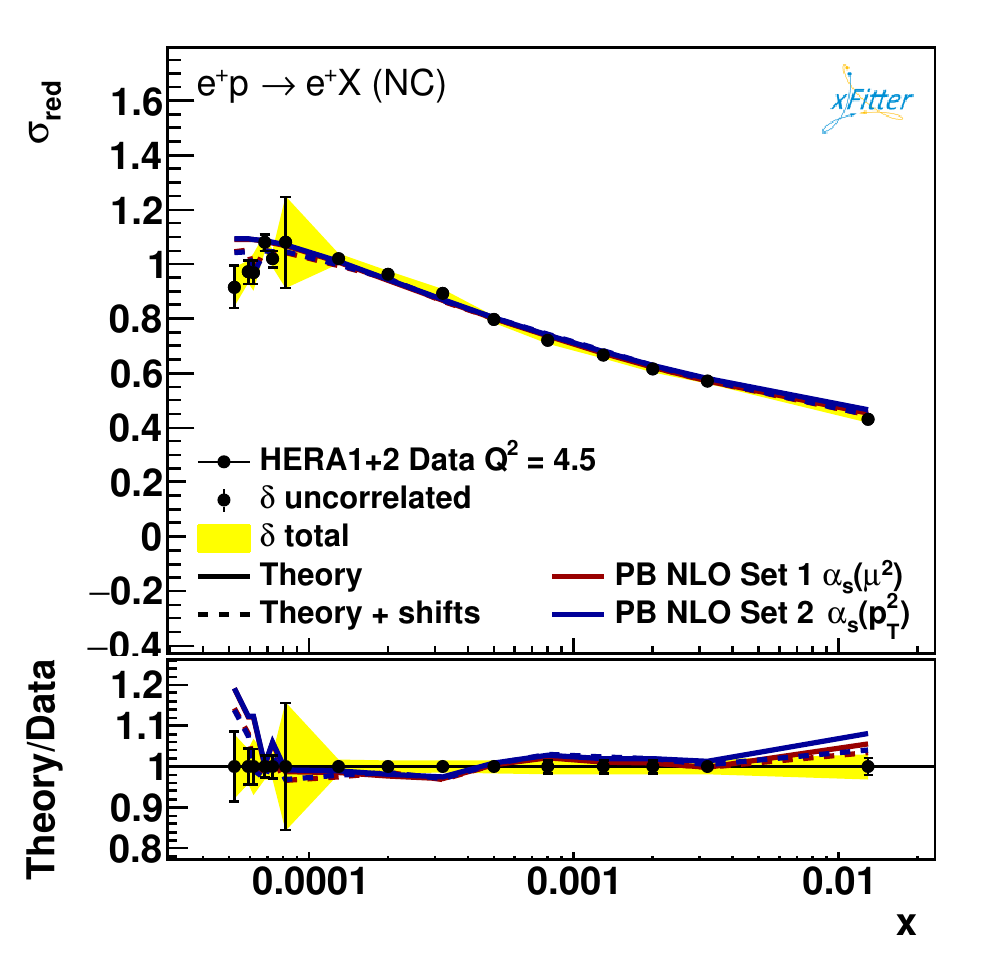}  
\includegraphics[width=0.3\textwidth]{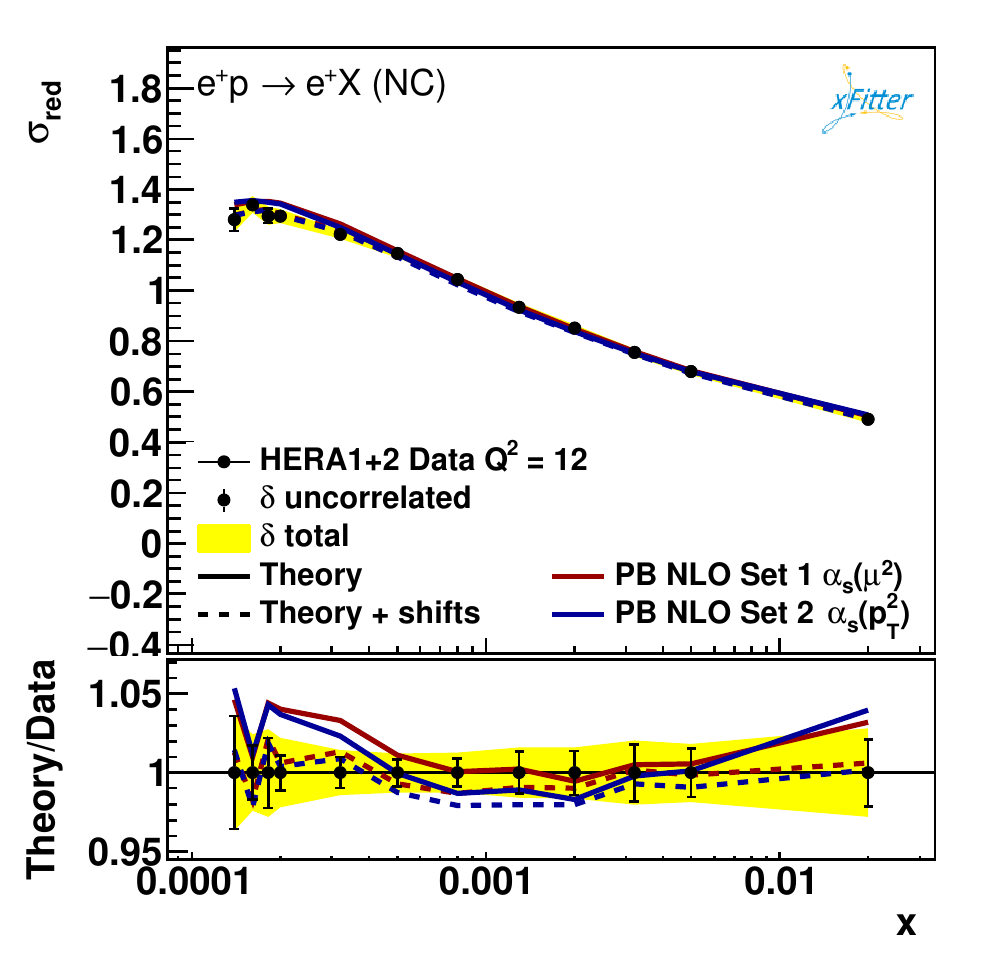}  
\includegraphics[width=0.3\textwidth]{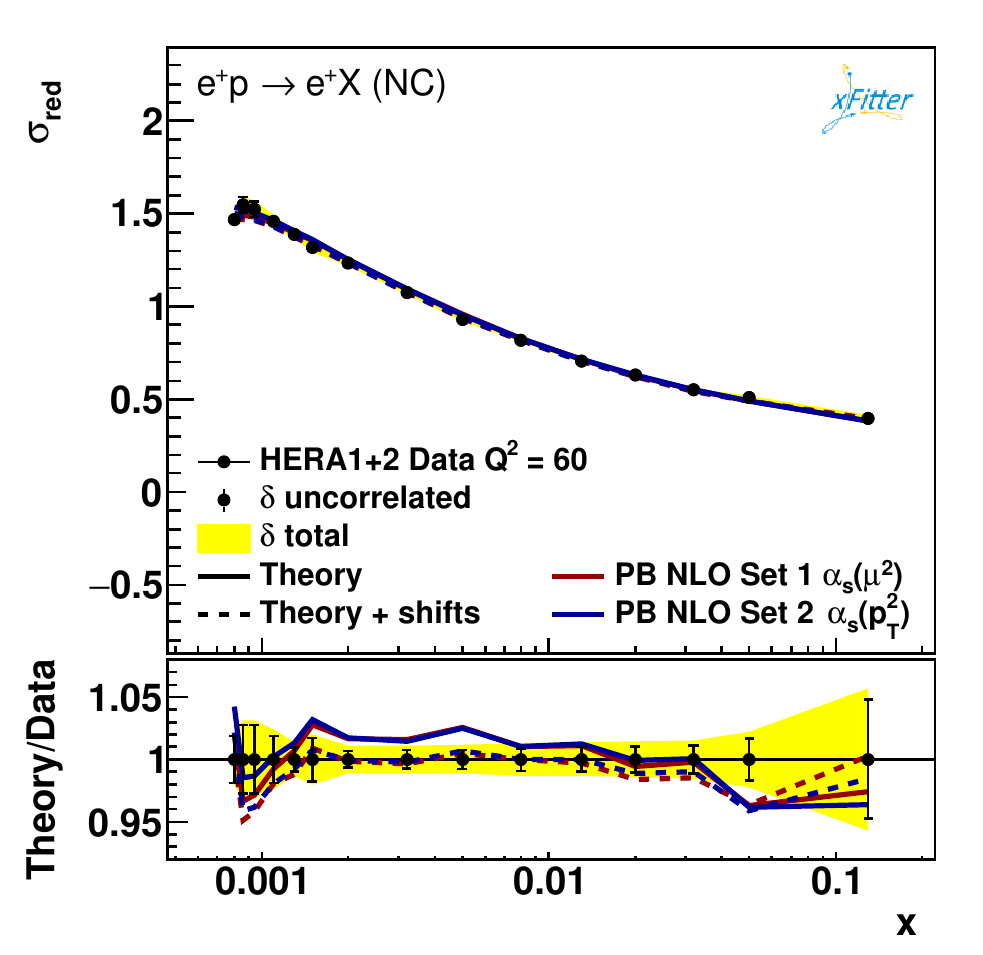}  
  \caption{\small Measurements of the reduced cross section~\protect\cite{Abramowicz:2015mha}   
  compared to predictions using PB - TMD Set~1 and Set~2 from~\protect\cite{Martinez:2018jxt}. 
}
\label{f2charm}
\end{center}
\end{figure}

\begin{figure}[htb]
\begin{center} 
\includegraphics[width=0.405\textwidth]{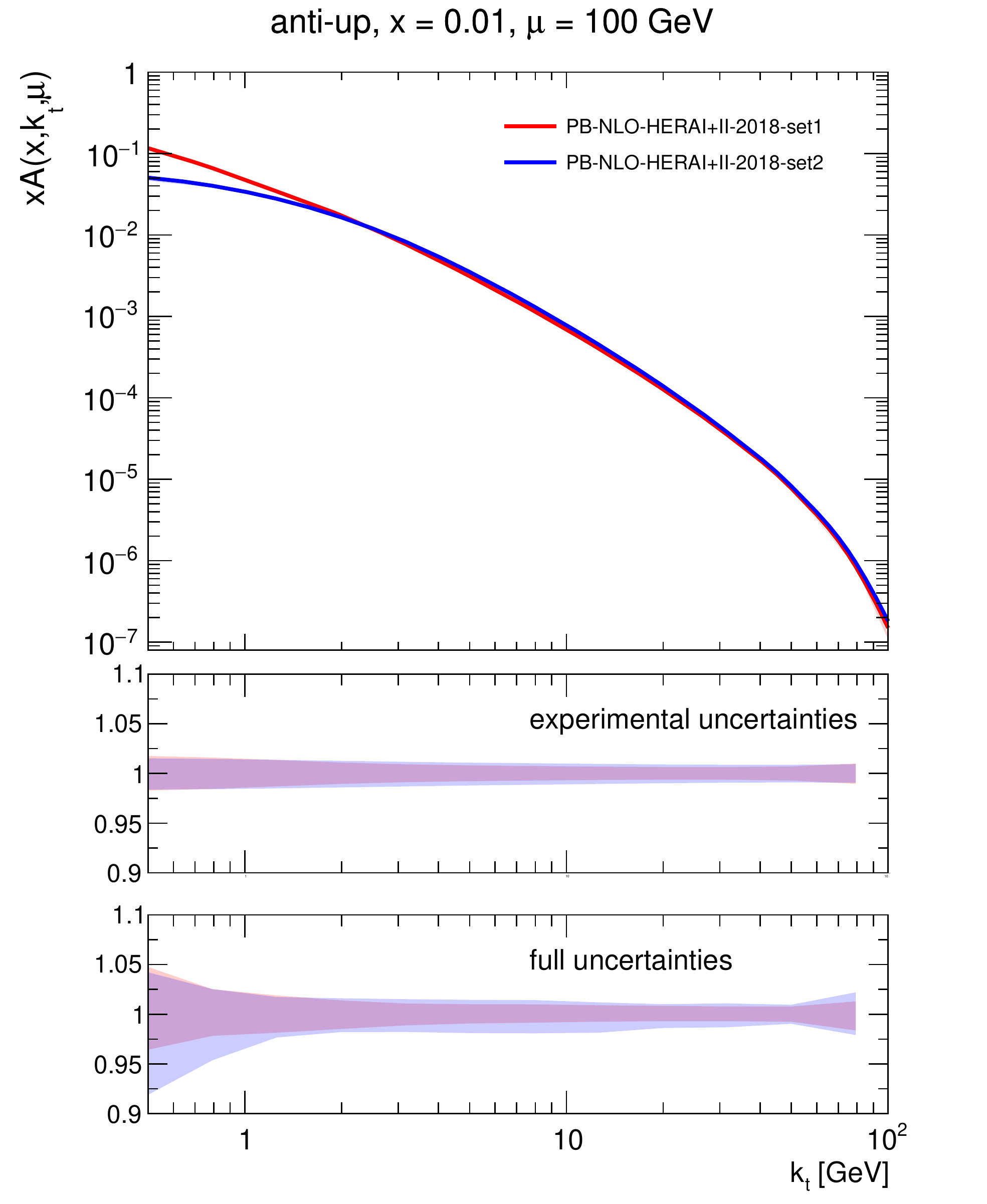}
\includegraphics[width=0.405\textwidth]{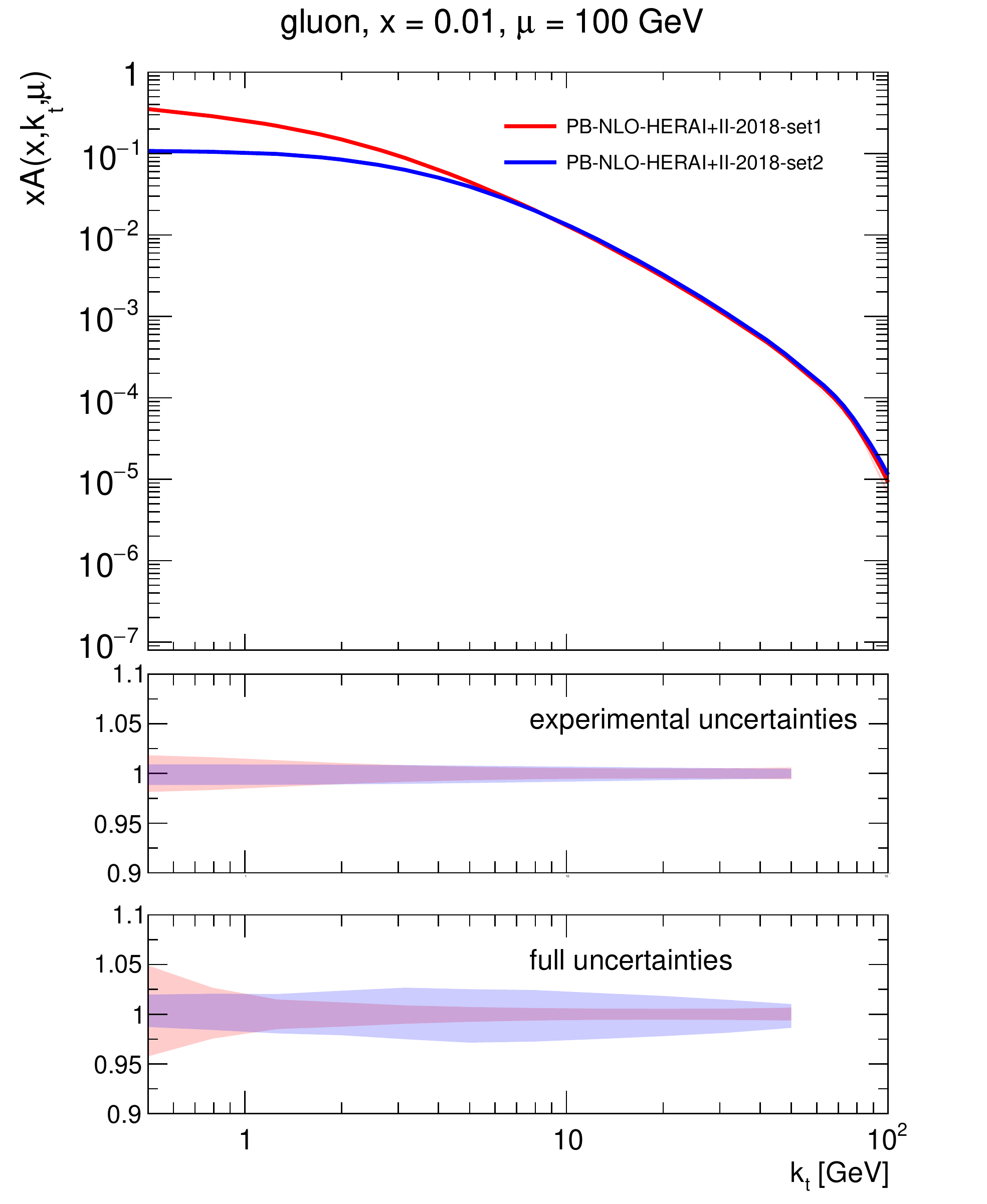}
  \caption{\small TMD   $\bar u$ and gluon  distributions as a function of $k_T$ for  $\mu=100$~GeV at $x=0.01$~\protect\cite{Martinez:2018jxt}. In the lower panels 
we show the relative uncertainties coming  from experimental uncertainties and the total of experimental and model uncertainties.
  }
\label{TMD_pdfs3}
\end{center}
\end{figure} 

\begin{figure}[htb]
\begin{center} 
\includegraphics[width=0.405\textwidth]{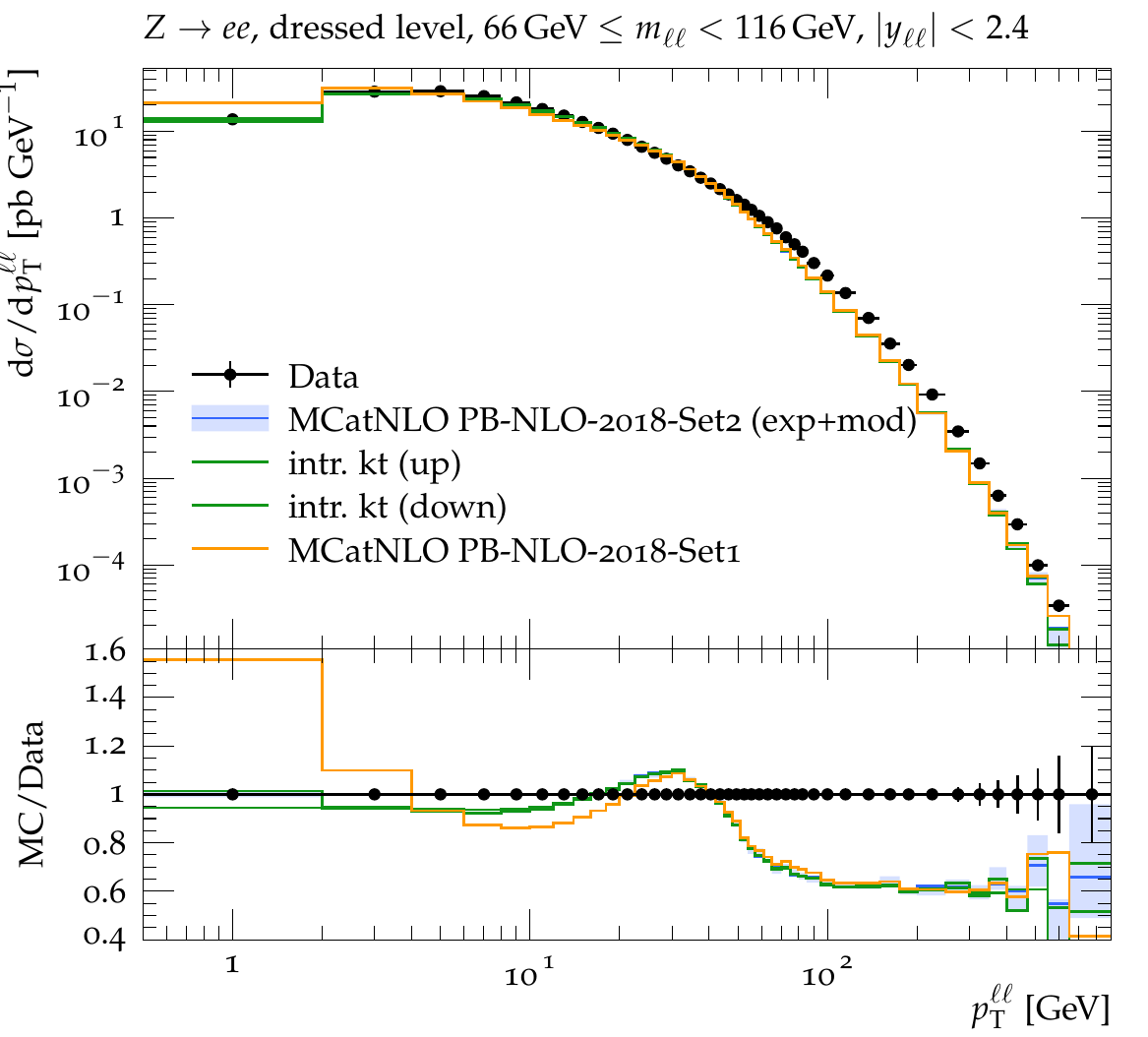} 
\includegraphics[width=0.405\textwidth]{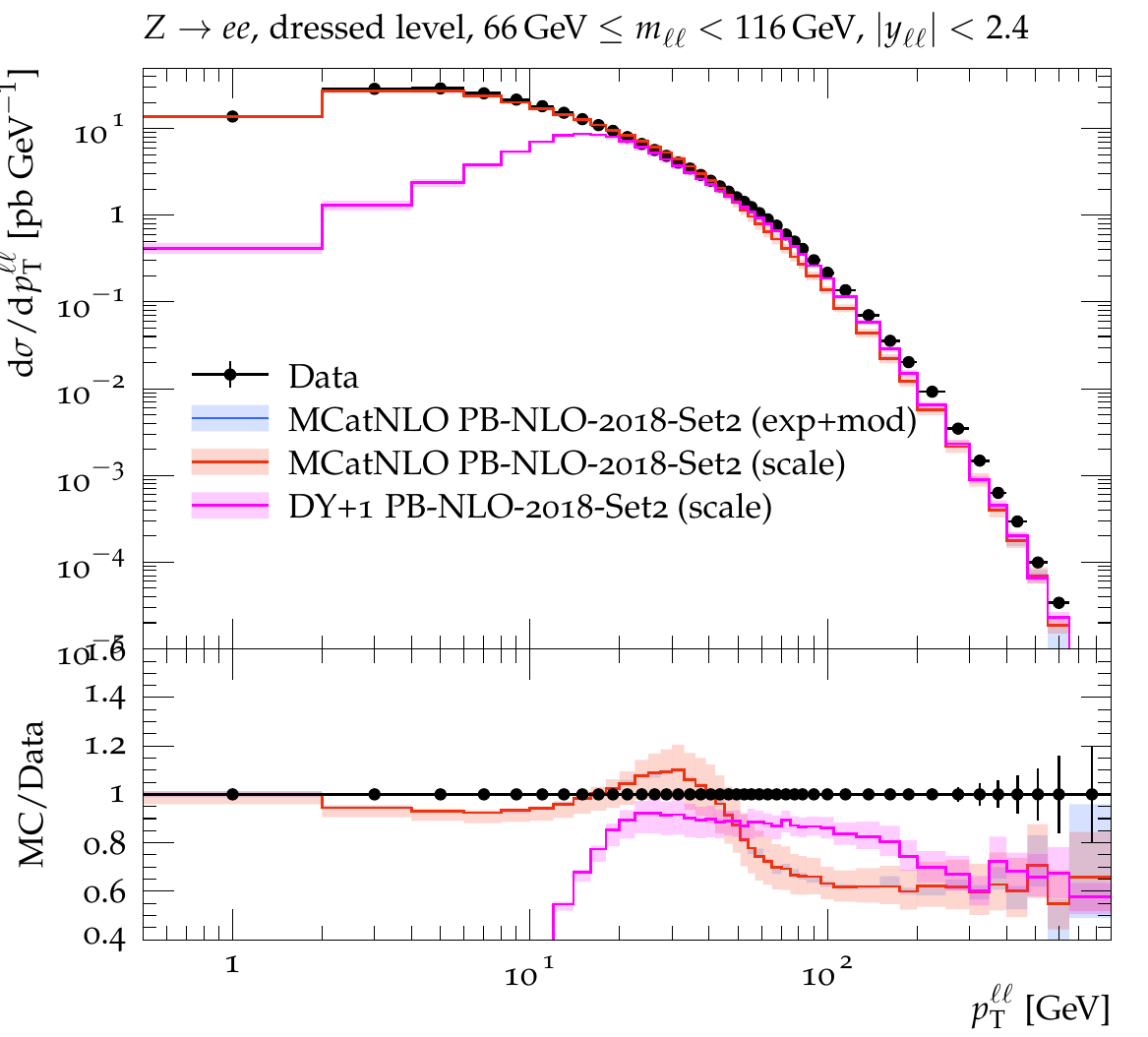} 
\caption{\small Transverse momentum $p_T$ spectrum of Z -bosons as measured by \protect\cite{Aad:2015auj} at $\sqrt{s}=8 $ TeV compared to the 
prediction~\protect\cite{Martinez:2019mwt}  using aMC@NLO and NLO PB -TMD. 
Left: uncertainties from the PB -TMD and 
 from changing the width of the intrinsic gaussian distribution by a factor of two. Right: with uncertainties from the TMDs  and scale variation combined. 
  }
\label{Zpt-TMD_uncertainty} 
\end{center}
\end{figure}

Figs.~\ref{f2charm}, \ref{TMD_pdfs3}~\cite{Martinez:2018jxt}  show results from PB fits  to the HERA high-precision inclusive 
DIS data~\cite{Abramowicz:2015mha}. The fits use NLO evolution kernels in Eq.~(\ref{integeqforA}) and NLO 
DIS hard-scattering coefficient functions~\cite{Botje:2010ay}. They    are performed 
using the  open-source fitting platform   \verb+xFitter+~\cite{Alekhin:2014irh} and the numerical techniques  
developed in~\cite{Hautmann:2014uua} to treat the transverse momentum dependence in the fitting procedure. 
Fig.~\ref{f2charm} illustrates the description of the reduced DIS cross section measured at HERA 
by the fitted TMDs.  In  Fig.~\ref{f2charm} two fitted TMD sets are presented,  differing  by the 
treatment of the momentum scale in the  coupling $\as$, so that one can 
compare the effects of $\as$ evaluated at the transverse momentum scale  
prescribed by the angular-ordered branching~\cite{Catani:1990rr,Martinez:2018jxt} with $\as$ evaluated at the evolution scale.  
The TMDs are extracted including a determination of experimental and theoretical uncertainties. An example of such results is given in 
Fig.~\ref{TMD_pdfs3}, showing the transverse momentum  dependence of $\bar u$ and gluon 
 distributions for fixed values of $x$ and $\mu$, and associated uncertainties. 

The  $k_T$ dependence in Fig.~\ref{TMD_pdfs3} results from  intrinsic transverse momentum and  evolution.   
The intrinsic  $k_T$ in  Fig.~\ref{TMD_pdfs3} is described by a simple gaussian at $\mu_0 \sim {\cal O}$ (1 GeV)  
with (flavor-independent and $x$-independent) width $\sigma = k_0 / \sqrt{2}$,  $k_0 = 0.5$ GeV. 
This is to be compared with higher values of intrinsic $k_T \sim$ 2 - 3 GeV obtained  from tuning in shower MC event  
generators~\cite{Bahr:2008pv,Sjostrand:2014zea}.

\begin{figure}[htb]
\begin{center} 
\includegraphics[width=0.495\textwidth]{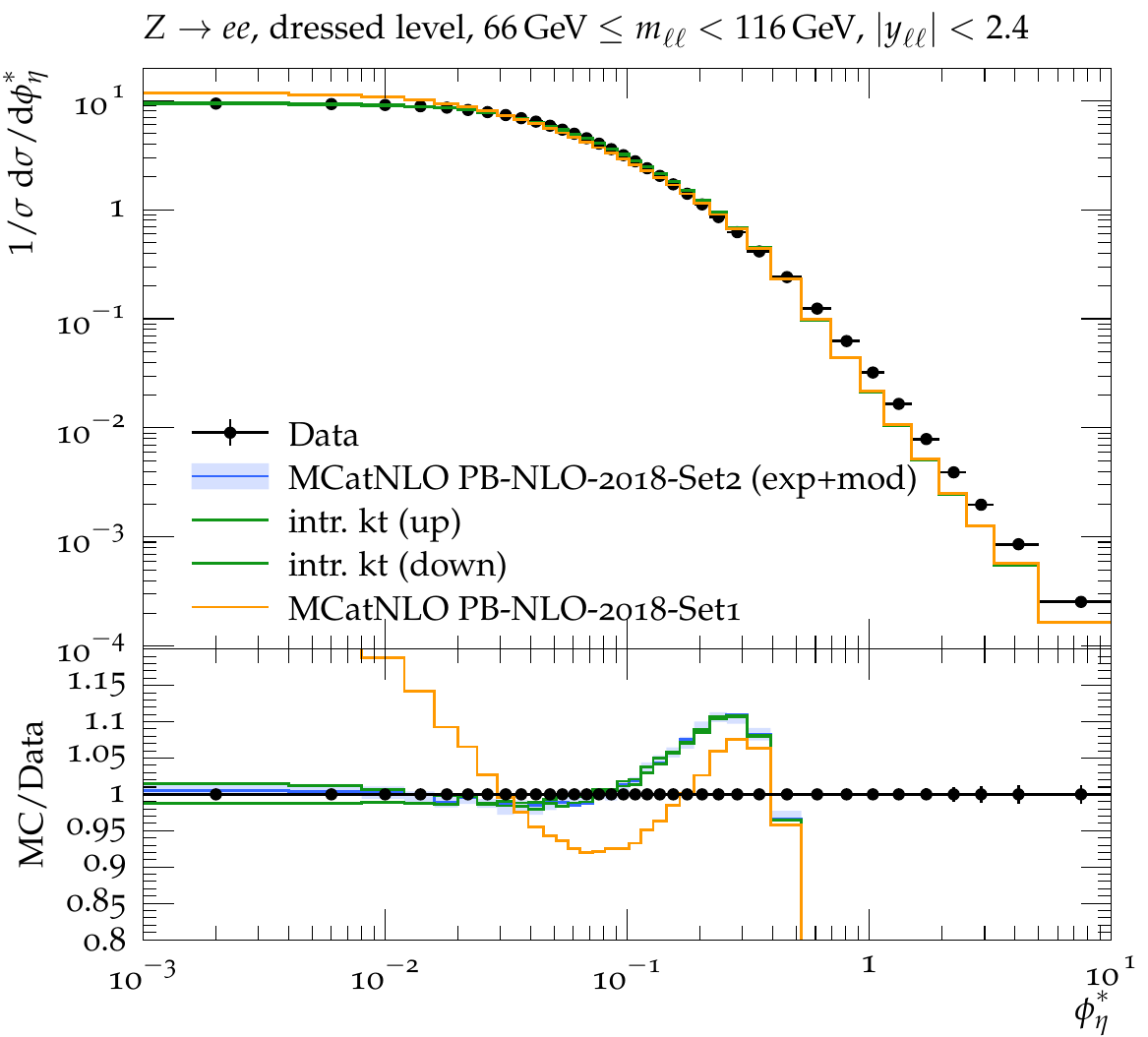} 
\includegraphics[width=0.495\textwidth]{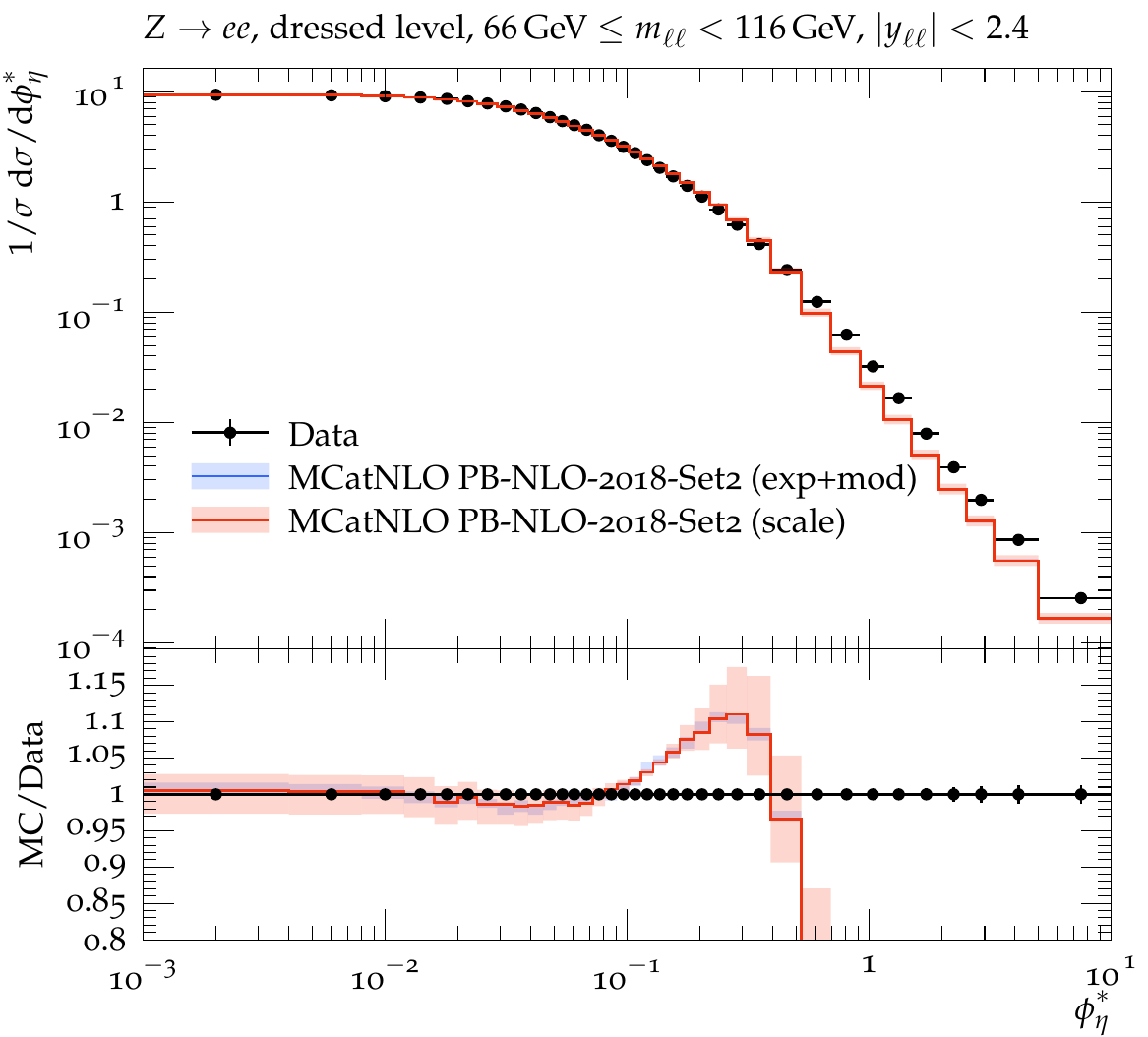} 
\caption{\small  $\phi^*$ spectrum of Z -bosons as measured by \protect\cite{Aad:2015auj} at $\sqrt{s}=8 $ TeV compared to the 
prediction~\protect\cite{Martinez:2019mwt} using  
aMC@NLO and NLO PB -TMD. 
Left: uncertainties from the PB -TMD and 
 from changing the width of the intrinsic gaussian distribution by a factor of two. Right: with uncertainties from the TMDs  and scale variation combined. 
  }
\label{Zphi-TMD_uncertainty} 
\end{center}
\end{figure}

In Figs.~\ref{Zpt-TMD_uncertainty}, \ref{Zphi-TMD_uncertainty}~\cite{Martinez:2019mwt}   the PB  TMDs are combined with the 
NLO calculation of 
 DY  $Z$-boson production to determine predictions  for  the lepton pair transverse momentum $p_T$  spectrum and $\phi^*$ spectrum and compare them with 
  LHC measurements~\cite{Aad:2015auj}.  
This computation requires addressing  issues of matching~\cite{jcc-fh-jhep}, analogous to those that arise in the case of parton showers. The 
matching is accomplished using the aMC@NLO framework~\cite{alwall14}, as described in~\cite{Martinez:2019mwt}.   
The calculations are performed using 
 {\sc Cascade}~\cite{Jung:2010si}   to read LHE~\cite{Alwall:2006yp}  files and produce output files,  
 and {\sc Rivet}~\cite{Buckley:2010ar} to analyze the outputs.

We see from the left panel in Fig.~\ref{Zpt-TMD_uncertainty}  that the spectrum at low $p_T$ is sensitive to the 
angular ordering effects embodied in the  different treatment of $\as$ in the  PB Set 1 and Set 2. 
The behaviors in the DY spectrum come from the  $k_T$  distributions in  Fig.~\ref{TMD_pdfs3}. 
The uncertainties on the DY predictions in Figs.~\ref{Zpt-TMD_uncertainty} and  \ref{Zphi-TMD_uncertainty}    
come from TMD uncertainites and scale variations, with the latter dominating the overall uncertainty. 
The bump in the $p_T$ distribution for intermediate $p_T$ is an effect of  the matching and 
the matching scale   --- a similar effect is seen when using parton showers instead of PB  TMD. 
The deviation in  the spectrum at higher $p_T$ is due to including only  ${\cal O} (\as) $ corrections but missing higher orders. We see from 
 the right panel of Fig.~\ref{Zpt-TMD_uncertainty}  that the contribution from DY + 1 jet at NLO plays an important role at 
 larger $p_T$.

\begin{figure}[htb]
\begin{center} 
\includegraphics[width=0.405\textwidth]{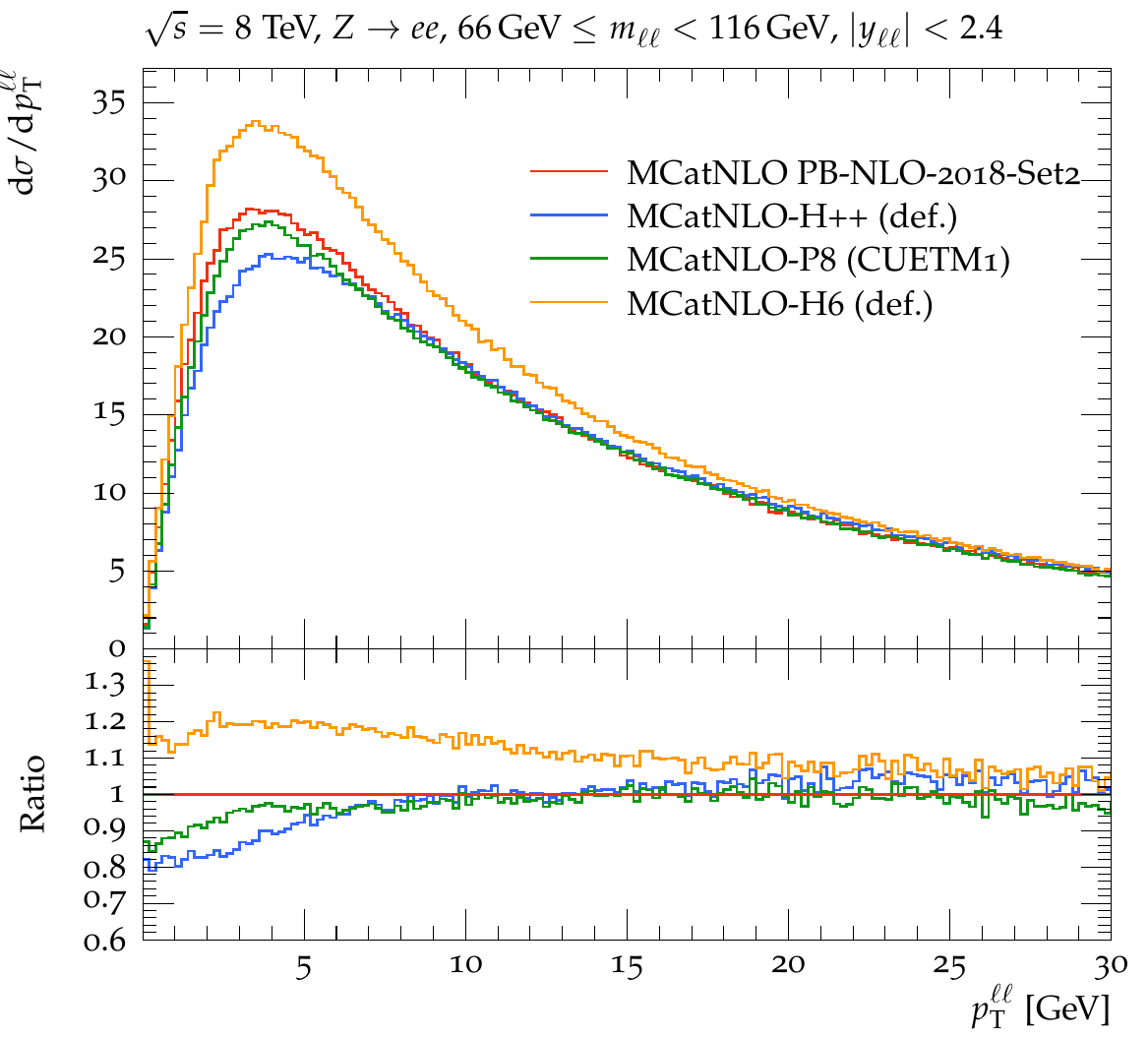} 
\includegraphics[width=0.405\textwidth]{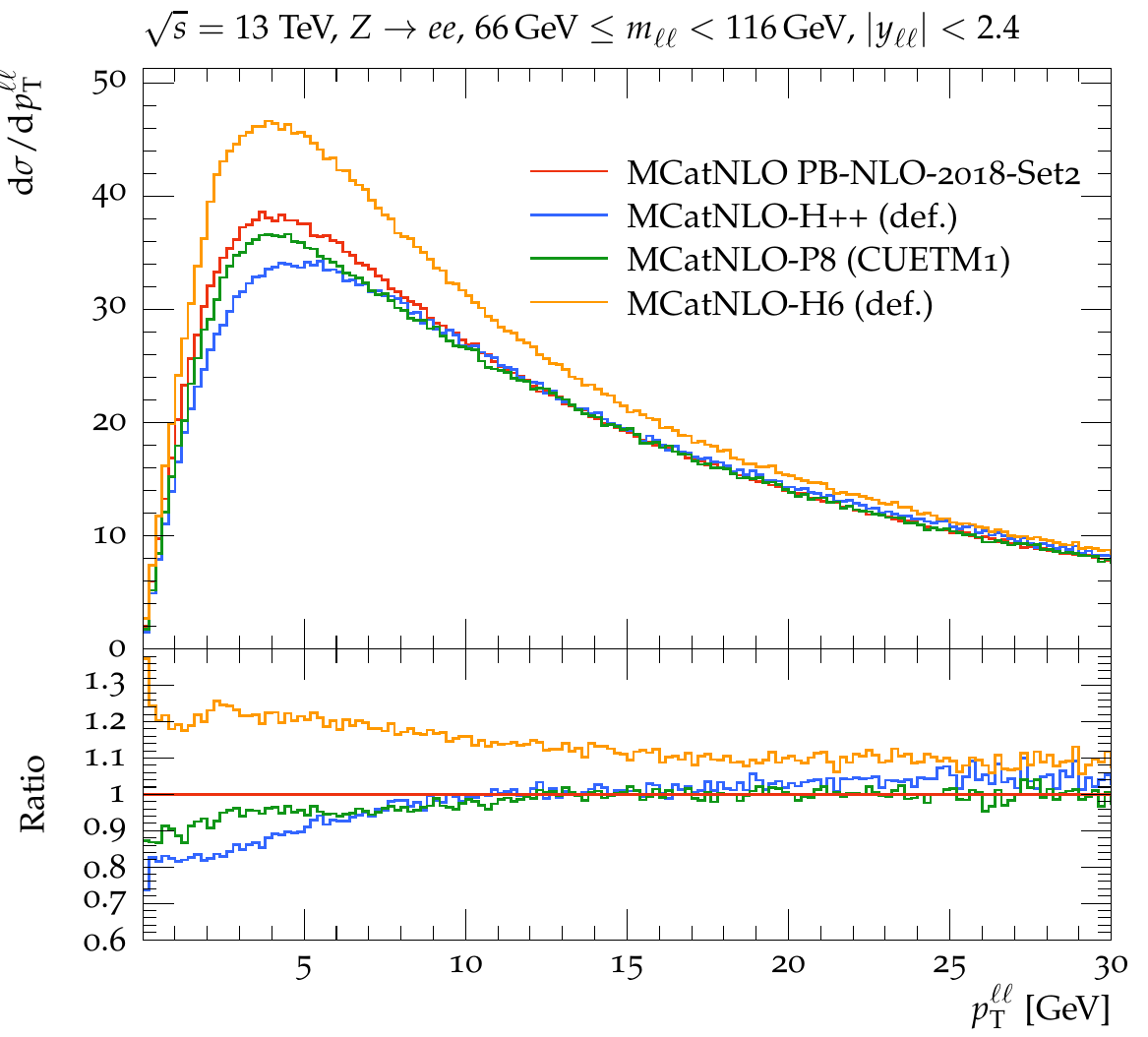} 
\caption{\small Transverse momentum $p_T^{ll}$ spectrum of Z -bosons at $\sqrt{s} = 8$ TeV(left) and $13$ TeV (right) obtained 
with the PB  method~\protect\cite{Martinez:2019mwt}, the parton shower of {\sc Pythia}8~\protect\cite{Sjostrand:2014zea} with 
tune CUETP8M1~\protect\cite{Khachatryan:2015pea},
{\sc Herwig}++~\protect\cite{Bahr:2008pv}, and {\sc Herwig}6~\protect\cite{Corcella:2002jc}.  }
\label{Zpt-FineBin} 
\end{center}
\end{figure}

In Fig.~\ref{Zpt-FineBin} we focus on the region of lowest transverse momenta accessible at the LHC, which is the region 
most sensitive to the nonperturbative and resummed QCD contributions.  We show predictions from PB  TMDs and from 
parton showers, all   of which are  obtained 
using the same aMC@NLO framework    for the hard process  at NLO (with appropriate subtractions terms, and the same collinear densities).  
While all calculations tend to agree for larger  $p_T$, differences are observed for  $p_T  $  < 5 - 10 GeV.  
In particular, the prediction using  {\sc Herwig}6, which  has parameter settings that  were not tuned to recent measurements,  serves 
as an illustration of the sensitivity of MC tunes.   Dedicated measurements with  fine   $p_T  $ binning in the region $p_T  $  <  5 - 10 GeV will 
allow one to investigate  TMD dynamics  and  analyze  resummation, showering and nonperturbative contributions.

\section{Conclusion}

MC event generators  
incorporating the dynamics of TMD parton distribution and 
fragmentation  functions are  instrumental in the development of high-energy 
physics programs which rely on precision experiments in hadron  
collisions at the highest luminosities   or highest energies. 

We have discussed TMD evolution in the parton branching formalism,   and 
 presented the TMD branching equation. 
 This approach is designed to be implementable in MC event generators, applicable 
 over  kinematic regions ranging  from low to high transverse momenta, connectible 
in a direct manner with  DGLAP evolution of collinear distributions. 

We have illustrated  ongoing  MC work, 
presenting results of matching PB TMDs with NLO calculations of DY lepton pair 
production via the method of aMC@NLO, and comparing the predictions thus obtained 
with measurements of DY $p_T$ and  $\phi^*$ spectra at the LHC. 

The PB approach can be extended in both its  nonperturbative and perturbative aspects.  On one hand, 
non-gaussian intrinsic distributions may be taken into account, including flavor-dependence and $x$-dependence. 
 This is relevant  to   perform  PB TMD fits  to experimental data for broader sets of processes 
 and observables besides inclusive DIS. 

On the other hand,   the logarithmic accuracy at low 
transverse momenta may be improved  through  perturbative coefficients in the Sudakov form factor and kernel 
of the TMD branching equation, and the finite-order accuracy at high transverse momenta  through 
matching with corrections  of  higher jet multiplicity and higher perturbative order.  

We have discussed PB TMDs in the unpolarized case. The extension to polarized TMDs in Table~1  
will be  relevant both for 
experiments with polarized hadron beams and for 
 studies of parton polarization effects, e.g.~double spin flip effects  for  gluon fusion processes,  in experiments with 
  unpolarized hadron beams. \\ 
\noindent 
{\bf Acknowledgments.}  I thank  the organizers   for the invitation  and  for the   pleasant   atmosphere at the Symposium.

\end{document}